\documentclass{ws-procs975x65}

\usepackage{amsfonts,amssymb,bbm}

\begin{document}
\providecommand{\1}{\ensuremath{\mathbbm{1}}}
\providecommand{\com}[1]{{\bf \ $\star $ #1 $\star$\ }}



\title{Averaged Energy Inequalities for Non-Minimally Coupled Classical Scalar Fields}

\author{Lutz W. Osterbrink}

\address{Department of Mathematics,\\
University of York, \\
Heslington, \\
York YO10 5DD, \\
United Kingdom,
\email{lwo500@york.ac.uk}}


\begin{abstract}
The stress-energy tensor for the non-minimally coupled scalar field is known not to satisfy the pointwise energy conditions, even on the classical level.
We show, however, that local averages of the classical stress-energy tensor satisfy certain inequalities and give bounds for averages along causal geodesics. It is shown that in vacuum background spacetimes, ANEC and AWEC are satisfied.
Furthermore we use our result to show that in the classical situation we have an analogue to the so called quantum interest conjecture. These results lay the foundations for averaged energy inequalities for the quantised non-minimally coupled fields.
\end{abstract}

\bodymatter

\section{Introduction}
It is generally believed that the energy density should be positive for all physically reasonable classical matter.
However, it is well known that this is not true for quantised fields.
Wightman fields, for example, do not satisfy pointwise positivity of the renormalised energy density \cite{EGJ65}, which resulted in a lot of research on this peculiarity. In particular the work of L.H.~Ford \cite{F78} was seminal and resulted in what is usually referred to as the {\it quantum inequalities}\footnote{For a good overview see, e.g., the work by C.J. Fewster\cite{F03} and references therein.}. They state that, even though the energy density (for instance) can be made arbitrarily negative at a point by varying the quantum states, the weighted time-like average is bounded from below. This bound is in particular state-independent.

Additionally to the violation on the quantum level, it is well known that the pointwise energy conditions can even be violated on the classical level.
One of the theories allowing such violations is the classical scalar field, non-minimally coupled to the Ricci-scalar of the spacetime manifold. Such a coupling changes the form of the energy density, even in the limit of a flat spacetime, such that the pointwise energy conditions can be violated. This can actually be so severe that it is possible to find wormhole spacetimes\cite{BV99,BV00}, supported by the non-minimally coupled scalar field.
On the other hand, there are various reasons to believe that such effects should be limited by certain bounds to the energy density, at least its weighted averages. One of those reasons, and probably the most obvious, is that there must be restrictions such that the second law of thermodynamics is not violated, at least on a macroscopic\footnote{The parameter defining macroscopic in the classical field theory is the maximal field amplitude.} scale. In particular, this means that there must be limitations (of some kind) to the duration and amplitude of the negative energy density. These should then rule out any possibility to use negative energy density to cool down a hot body without (macroscopically) changing its entropy \cite{F78}. 

Below, we give an overview of the work done so far, to find such restrictions for the classical scalar field with non-minimal coupling, based on the results obtained by the author together with C.J.~Fewster\cite{FO06,FO06b}.

\section{Bounds for the Classical Non-Minimally Coupled Scalar Field}
The stress-energy tensor for the non-minimally coupled scalar field can be derived from its Lagrangian,
$L=\frac{1}{2}\left\{ (\nabla \phi)^2 -(m^2+\xi R)\phi^2 \right\}$, 
by variation of the action with respect to the co-metric $g^{\mu\nu}$. 
A straightforward calculation yields the expression\footnote{See [\citen{FO06}] for conventions.}
\begin{equation}\label{eq_st0}
T_{\mu \nu}=\left(\nabla_\mu \phi \right)\left( \nabla_\nu \phi\right) +\frac{1}{2}g_{\mu\nu} \left(m^2\phi^{2}-(\nabla \phi)^2\right)+\xi \left\{ g_{\mu\nu}\square_{g} -\nabla_\mu \nabla_\nu-G_{\mu\nu}\right\} \phi^2,
\end{equation}
where $G_{\mu\nu}$ is the Einstein tensor and $\square_{g}$ is the d'Alembertian with respect to the metric $g$. Furthermore, the equation of motion is $(\square_{g}+m^{2}+\xi R)\phi =0$.
Even though the Lagrangean and the equation of motion in flat spacetime reduce to the one for minimal coupling, i.e., for $\xi=0$, the stress-energy tensor (\ref{eq_st0}) does not.
This feature makes it possible to have negative energy density for the non-minimally coupled scalar field, even in flat spacetimes.
A simple example is given by L.H.~Ford and T.A.~Roman in [\citen{FR01}].

The averaged stress-energy tensor, however, obeys the following result \cite{FO06}:
\begin{theorem}\label{thm_classgeod}
Let $\gamma$ be a causal geodesic with affine parameter $\lambda$ in a spacetime $(M,g)$.
Furthermore, let $T_{\mu\nu}$ be the stress-energy tensor of the non-minimally coupled classical scalar field with
coupling constant $\xi\in [0,1/4]$. For every
 real-valued function $f\in \mathcal{C}^{2}_{0}(\mathbb{R})$ the inequality
\begin{equation*}\label{eqn_geodineq}
\int_{\gamma} \mathrm{d} \lambda \ T_{\mu\nu} \dot{\gamma}^\mu \dot{\gamma}^\nu f^2
\geq -2\xi\int_{\gamma} \mathrm{d} \lambda \ \left\{(\partial_{\lambda} f)^{2}+\frac{1}{2} R_{\mu\nu}\dot{\gamma}^\mu \dot{\gamma}^\nu f^2 -(\frac{1}{4}-\xi)R \dot{\gamma}^2f^{2}\right\}\phi^{2} 
\end{equation*}
is satisfied on-shell.
\end{theorem}
Here, ``on-shell'' means, that the field is required to satisfy the field equation, as given above.
This result can be used in various ways to analyse averaged energy densities and can be generalised to spacetime-volume averages\cite{FO06}. Interesting results for Ricci-flat spacetimes can be derived by scaling arguments. Without going into too much detail, we can summarise the results by: {\it Long-lasting negative energy densities of large magnitude must be associated with large magnitudes of the field or with large curvatures}. As a consequence, one finds conditions that ensure ANEC and AWEC. 

A further interesting aspect of our work concerns {\it energy interest}. Originally analysed in quantum field theory, this phenomenon was first described by Ford and Roman \cite{FR99}. It states that negative energy density is always associated with positive energy density, which actually overcompensates the former one, ensuring an overall positive energy density. This overcompensation can then be understood metaphorically as the repayment with interest of a negative energy density debt. 
The same phenomenon can be found for the classical non-minimally coupled scalar field\cite{FO06}. In detail, one finds that the maximal time-separation of such pulses is proportional to the coupling constant, the maximal field amplitude and furthermore inversely proportional to the magnitude of the negative energy density.

Since the non-minimally coupled scalar field allows these strange phenomena already on the classical level, it is very important to study them for the quantised field as well. 
To get a lower bound for the latter situation one has to mix two different methods.
One of these is analogous to the classical manipulation described above and the other is in line with
the methods used by Fewster and Eveson\cite{FE98} to derive a class of quantum inequalities.
As expected, their result is recovered in the case of minimal coupling.
The more general result that we found\cite{FO06b} is a lower bound for the time-like averaged energy density $\hat{\rho}_{f}$ with coupling constants $\xi\in [0,1/4]$. It is given by
\begin{equation}\label{quant_ineq}
\hat{\rho}_{f}\geq - (1-4\xi)\ \mathfrak{Q}^\xi_{FE}(f)\1-2\xi\ \hat{\mathfrak{B}}(f),
\end{equation}
in terms of quadratic forms.
The non-linear functional $\mathfrak{Q}^{\xi=0}_{FE}(f)$ is the one that was obtained as the state independent lower bound for the minimal coupling, as remarked above. The additional term $\hat{\mathfrak{B}}(f)$ is a non-negative quadratic form, whose expectation values are state-dependent. Even though one can show that the right hand side in (\ref{quant_ineq}) is unbounded from below, there is a sense in which the bound is nontrivial, in that
$\hat{\mathfrak{B}}(f)$ is of ``lower order'' than the energy density. Our hope is that by understanding this case, we will be better placed to understand quantum energy inequalities for general interacting quantum fields.


\vfill

\end{document}